\documentclass[final,5p]{elsarticle}

\usepackage{amssymb}
\usepackage{bbm}
\usepackage{amssymb,graphicx}
\usepackage[intlimits]{amsmath}
\usepackage[small]{subfigure}
\usepackage{color}
\usepackage{array}
\usepackage{multirow}

\usepackage[colorlinks=true]{hyperref}
\hypersetup{
 linkcolor=blue,citecolor=blue,urlcolor=black}

\journal{Physics Letters B}

\begin{document}

\begin{frontmatter}



\title{Overlaps for Matrix Product States of Arbitrary Bond Dimension in ABJM theory}


\author[bdt,wig]{T.\ Gombor}
\author[cph]{C.\ Kristjansen}

            \affiliation[bdt]{organization={Department of Theoretical Physics, E\"{o}tv\"{o}s Lor\'{a}nd Univeristy },
            addressline={P\'{a}zm\'{a}ny P\'{e}ter stny.\ 1A}, 
            city={Budapest},
            postcode={1117}, 
            country={Hungary}}
            
\affiliation[wig]{organization={Wigner Research Centre for Physics},
            addressline={Konkoly-Thege Miklós út 29-33.}, 
            city={Budapest},
            postcode={1121}, 
            country={Hungary}}

\affiliation[cph]{organization={Niels Bohr Institute, Copenhagen University},
            addressline={Blegdamsvej 17}, 
            city={Copenhagen},
            postcode={2100}, 
            country={Denmark}}

\begin{abstract}
We find a closed formula for the overlap of Bethe eigenstates of an alternating $SU(4)$ spin chain, describing the scalar sector of ABJM theory, and matrix product states of
any bond dimension representing 1/2 BPS co-dimension one domain walls in the field theory. One point functions of the defect CFTs involved, being directly expressible in terms of these
overlaps, are hence completely determined. 
\end{abstract}






\end{frontmatter}


\section{Introduction}
Matrix product states, a tool from quantum information theory, were put to use in holography in the computation of one-point functions in defect CFTs being domain wall
versions of ${\cal N}=4$ SYM theory~\cite{deLeeuw:2015hxa,Buhl-Mortensen:2015gfd}. The matrix product states could be of any bond dimension, $d$, and the matrices involved were
matrices generating an irreducible $d$-dimensional representation of either $SU(2)$, $SU(2)\times SU(2)$  or $SO(5)$ where the dimension of the representation was related to a jump in the rank of the gauge group of the underlying field theory across the domain wall~\cite{Constable:1999ac,Constable:2001ag,Kristjansen:2012tn}.  One-point functions of conformal operators being given by the overlap between the matrix product state and  Bethe eigenstates of the integrable spin chain underlying ${\cal N}=4$ SYM~\cite{Beisert:2010jr} turned out to be computable in closed form in many cases~\cite{deLeeuw:2015hxa,Buhl-Mortensen:2015gfd,deLeeuw:2016umh,DeLeeuw:2018cal,DeLeeuw:2019ohp,Komatsu:2020sup,Gombor:2020kgu,Gombor:2020auk}
and entailed new developments in the understanding of boundary integrability from the statistical mechanical point of view~\cite{Piroli:2017sei,Pozsgay:2018dzs,}. 
Furthermore, certain 
matrix product states were understood to encode, via overlaps,  the correlation function between giant gravitons and a single trace operator in ${\cal N}=4$ SYM, and exact results were found 
for the case of two giant gravitons which corresponds to a matrix product state of bond dimension two~\cite{Jiang:2019zig,Jiang:2019xdz}.

ABJM theory, while being a close cousin of ${\cal N}=4$ SYM as the field theory occurring in a lower dimensional version of the AdS/CFT correspondence~\cite{Aharony:2008ug}, symmetry wise differs from the latter by being less than maximally supersymmetric, having its fields in a bi-fundamental rather than adjoint representation of the gauge group and by having an unknown interpolating function in its integrability description~\cite{Klose:2010ki}.  In particular, the integrable spin chain encoding the spectral problem of  the theory is a chain where the representation alters between even and odd sites. The investigation of  boundary integrability of ABJM theory and its associated spin chain via the use of matrix product states is much less developed
than for ${\cal N}=4$ SYM and is likely to provide us with  novel examples of integrable boundary states and further insight on the interplay between defects and integrability.  The
program of calculating correlation functions between giant gravitons and single trace operators by means of matrix product states has been set up in~\cite{Yang:2021hrl}, see also~\cite{Chen:2019kgc} and exact results have been presented for the case of two giant gravitons \cite{Yang:2021hrl}. ABJM theory also allows for a defect set-up with a 1/2 BPS co-dimension one 
domain wall~\cite{Terashima:2008sy,Chandrasekhar:2009ey} and scalar one-point functions in this set-up can be calculated using matrix product states of an alternating integrable $SU(4)$ spin chain~\cite{Kristjansen:2021abc}.  A derivation of the relevant overlap formula for
a matrix product state of bond dimension 2 was presented in~\cite{Gombor:2021hmj}. In the present paper we consider matrix product states of arbitrary bond dimension and find
a closed expression for the overlaps, thereby providing a whole new class of integrable matrix product states. We note that the ability of treating matrix product states of arbitrary bond dimension is useful for making connection with string theory computations, as the bond dimension translates via
the holographic dictionary to a string theory flux which can be taken large and allow for a semi-classical approach~\cite{Nagasaki:2012re,Kristjansen:2012tn,Linardopoulos:2021rfq,Linardopoulos:2022wol}.

 We begin in section~\ref{Model} by introducing the integrable alternating $SU(4)$~spin chain and the matrix product states representing the 1/2 BPS co-dimension one domain walls in ABJM theory.  Subsequently, in section~\ref{integrability} we show that these matrix product states constitute integrable
  boundary states according to the criterion of~\cite{Piroli:2017sei}.  We then, in section~\ref{overlap}, derive
 a closed overlap formula for any such matrix product state, i.e.\ in particular for any bond dimension,  by following a recursive strategy suggested in~\cite{Gombor:2020auk}. As this strategy builds on a number
of assumptions which can not immediately be verified we validate our result by a detailed numerical check in section~\ref{Validation}. Finally, section~\ref{Conclusion} contains our conclusion. 
 \section{The Model\label{Model}}
As it is the case for  $AdS_5/CFT_4$,  the planar spectral problem of the $AdS_4/CFT_3$ correspondence is solved by an integrable super spin chain~\cite{Minahan:2008hf,Gromov:2008qe}. In particular, at the two-loop level the conformal single trace scalar operators of the 3-dimensional field theory involved,  i.e.\ ABJM theory, constitute the eigenstates of an alternating $SU(4)$ spin chain
whose Hamiltonian reads~\cite{Minahan:2008hf}
\begin{eqnarray}\label{1-loop-mixing}
 H&=&\lambda ^2\sum_{l=1}^{2L}\left(1-P_{l,l+2}+\frac{1}{2}\,P_{l,l+2}K_{l,l+1} \right. \nonumber\\
&& \left. +\frac{1}{2}\,K_{l,l+1}P_{l,l+2}\right),
\end{eqnarray}
where $K_{l,m}$ and $P_{l,m}$ are the trace and permutation operators and where $\lambda$ is the 't Hooft coupling of
ABJM theory. 

The odd and even sites of the chain carry 
respectively the representations $\bf{4}$ and $\bf{\bar{4}}$ of $SU(4)$ corresponding to the following composition of the
single trace operators
\begin{equation}\label{spin-chain}
 \mathcal{O}=\Psi _{A_1\,\ldots\, A_{2L-1}}^{\hphantom{A}A_2\,\ldots\, A_{2L}}\mathop{\mathrm{tr}}Y^{A_1}Y^\dagger _{A_2}\ldots Y^{A_{2L-1}}Y^\dagger _{A_{2L}} \equiv |\Psi \rangle,
\end{equation}
with $A_i\in \{1,2,3,4\}$. 

The field theory has a classical solution where a subset of fields acquire a vacuum expectation value different from zero on one side (say $x>0$) of a co-dimension one domain wall~\cite{Terashima:2008sy},
a solution which can conveniently be described as a Nahm pole solution for  certain composite fields~\cite{Kristjansen:2021abc}
\begin{equation}
 \langle Y^\alpha Y^\dagger _\beta \rangle= \frac{M^{\alpha}_{\,\,\,\,\beta}}{x}, \hspace{0.5cm} x> 0, \hspace{0.5cm}
 \alpha,\beta \in \{1,2\}, \nonumber
 \end{equation}
where 
\begin{equation} \label{eq:M}
 M^{\alpha}_{\,\,\,\,\beta}=
 \frac{q}{2} \, \mathbbm{1}_{q-1} \otimes  (\sigma^0)^{\alpha}_{\,\,\,\,\beta}  + \, t_i^{(q-1)} \otimes (\sigma^i)^{\alpha}_{\,\,\,\,\beta}, 
\end{equation}
with the $t_i^{(q-1)}$ constituting a $q-1$ dimensional representation of $SU(2)$, i.e.
\begin{equation}
 [t^i,t^j]=i\varepsilon ^{ijk}t^k.
\end{equation}
These vevs break the $SU(4)$ symmetry of the scalar sub-sector to $SU(2)\times SU(2)\times U(1)$. The domain wall is
conveniently described as a boundary state in the form of a matrix product state of bond dimension $q-1$. More precisely,
$$ {\rm MPS} _{\hphantom{A}\beta_2\,\ldots\, \beta_{2L}}^{\alpha_1\,\ldots\, \alpha_{2L-1}}=\mathop{\mathrm{tr}}M^{\alpha_1} _{\,\,\,\,\,\beta_2}\ldots M^{\alpha_{2L-1}}_{\,\,\,\,\beta_{2L}},
$$
where the trace is over the $q-1$-dimensional auxiliary space.
The one-point functions of the field theory (for $x>0)$ in the presence of the domain wall can then be expressed as the following spin 
chain overlap~\cite{deLeeuw:2015hxa,Buhl-Mortensen:2015gfd}\footnote{We note that there exists an equally valid way
of expressing the one-point function in terms of a matrix product state, $\widehat{\mbox{MPS}}$, built from matrices encoding the vevs of  composite fields of the type $Y_{\beta}^{\dagger} Y^{\alpha}$~\cite{Kristjansen:2021abc}.}
\begin{equation}\label{O(x)}
 \left\langle \mathcal{O}(x)\right\rangle=\frac{1}{x^L}\,\,\frac{1}{\lambda ^LL^{\frac{1}{2}}}\,\,\frac{\left\langle {\rm MPS}_q\right.\!\left| \Psi \right\rangle}{\left\langle \Psi \right.\!\left|\Psi  \right\rangle^{\frac{1}{2}}}, \hspace{0.5cm} x>0,
\end{equation}
where we can write the matrix product state as, cf.~eqn~\eqref{spin-chain}
\begin{equation} \label{eq:MPS}
\langle \mbox{MPS}_q|= \mbox{tr} \,M^q_{1,2}M^q_{3,4}\ldots M^q_{2L-1, 2L},
\end{equation}
with the subscript $q$ referring to the representation of dimension $q-1$.
Following standard conventions we chose the pseudo vacuum of our spin chain to correspond to the operator
\begin{equation}\label{vac}
 \mathcal{O}_{\rm vac}=\mathop{\mathrm{tr}}(Y^1Y^\dagger _4)^L.
\end{equation}
As this state does not have any overlap with the matrix product state defined above, and our method for deriving the
overlap formula relies on this overlap being non-zero, we will perform a rotation of the matrices $M$ by a certain angle
$\theta$ that we will eventually send to zero.
First, we extend the definition of the matrices $ M^{\alpha}_{\,\,\,\,\beta}$ to $\alpha,\beta \in \{1,2,3,4\}$ by 
simply setting matrix elements not defined above to zero. Secondly, we define a rotated version of the matrix product state
in terms of  rotated versions of the $M$-matrices given by
\begin{equation}
M^{q,\theta}=g(\theta) M^q g(-\theta),
\end{equation}
where 
\begin{equation}
g(\theta)=
\begin{pmatrix}
\cos \theta & 0 & 0& -\sin \theta \\
0 & \cos \theta & -\sin \theta & 0 \\
0 & \sin \theta & \cos \theta& 0 \\
\sin \theta & 0 & 0& \cos \theta
\end{pmatrix}.
\end{equation}
We denote the corresponding rotated version of the matrix product state as $|\mbox{MPS}_{q,\theta}\rangle$.

\section{The integrability of the MPS \label{integrability}}
In this section we show that the MPS \eqref{eq:MPS} preserves ``half'' of the conserved charges, i.e.\ the MPS is integrable in the sense of \cite{Piroli:2017sei}.
 
The transfer matrices of the alternating $SU(4)$ spin chain can be obtained from the following $R$-matrix
\begin{equation}
R_{ab}(u)=u-P_{ab},
\end{equation}
and its crossed version, $\bar{R}_{ab}(u)=R_{ab}(2-u)^{t_{b}}$,
more precisely
\begin{align}
\mathcal{T}(u) & =\mathrm{Tr}_{a}R_{a1}(u)\bar{R}_{a2}(u)\dots R_{a,2L-1}(u)\bar{R}_{a,2L}(u),\\
\bar{\mathcal{T}}(u) & =\mathrm{Tr}_{a}\bar{R}_{a1}(u)R_{a2}(u)\dots\bar{R}_{a,2L-1}(u)R_{a,2L}(u).
\end{align}
These transfer matrices commute with each other
\begin{equation}
\left[\mathcal{T}(u),\mathcal{T}(u)\right]=\left[\mathcal{T}(u),\bar{\mathcal{T}}(u)\right]=\left[\bar{\mathcal{T}}(u),\bar{\mathcal{T}}(u)\right]=0,
\end{equation}
and they generate the Hamiltonian \eqref{1-loop-mixing}.

The $K$-matrix 
\begin{multline}
\mathbb{K}^{q,\gamma}(u) = \\
\left(\begin{array}{cccc}
\frac{a(u)}{2}+ut_{3}^{(q-1)} & u t_{-}^{(q-1)} & 0 & 0\\
u t_{+}^{(q-1)} & \frac{a(u)}{2}-ut_{3}^{(q-1)} & 0 & 0\\
0 & 0 & \frac{b(u)}{2} & 0\\
0 & 0 & 0 & \frac{b(u)}{2}
\end{array}\right)
\end{multline}
satisfies the following reflection equation
\begin{multline}
R_{ab}(u-v)\mathbb{K}_{a}^{q,\gamma}(u)R_{ab}(u+v)\mathbb{K}_{b}^{q,\gamma}(v) = \\
\mathbb{K}_{b}^{q,\gamma}(v)R_{ab}(u+v)\mathbb{K}_{a}^{q,\gamma}(u)R_{ab}(u-v)
\end{multline}
where $t_{\pm}^{(q-1)} = t_{1}^{(q-1)} \pm i t_{2}^{(q-1)}$ and
\begin{align}
a(u) & =-u^{2}+u+\gamma^{2}+\gamma-\frac{1}{4}q(q-2)\\
b(u) & =u^{2}-(2\gamma+1)u+\gamma^{2}+\gamma-\frac{1}{4}q(q-2)
\end{align}
Taking $\gamma=\frac{q}{2}$ the $K$-matrix at $u=1$ simplifies
as
\begin{equation} \nonumber
\mathbb{K}^{q,\frac{q}{2}}(1)=\left(\begin{array}{cccc}
\frac{q}{2}+t_{3}^{(q-1)} & t_{1}^{(q-1)}-it_{2}^{(q-1)} & 0 & 0\\
t_{1}^{(q-1)}+it_{2}^{(q-1)} & \frac{q}{2}-t_{3}^{(q-1)} & 0 & 0\\
0 & 0 & 0 & 0\\
0 & 0 & 0 & 0
\end{array}\right)
\end{equation}
which is just the matrix $M$ of eqn.\ \eqref{eq:M}. Using the argument of \cite{Pozsgay:2018dzs}, appropriately adapted to our case \cite{Gombor:2020kgu},  we obtain 
\begin{equation} \label{eq:intcond}
\langle\mathrm{MPS}_{q}|(\mathcal{T}(u)-\mathcal{T}(2-u))=
0,
\end{equation}
which proves the integrability of the MPS.

\section{The overlap formula \label{overlap}}

A closed formula for the overlap between the matrix product state and Bethe eigenstates with paired root configurations (to be specified below) is
expected to take the form
\begin{equation}
\frac{\langle\mathrm{MPS}_{q,\theta}|\mathbf{u}\rangle}{\sqrt{\langle\mathbf{u}|\mathbf{u}\rangle}}=\left[\sum_{k=1}^{q-1}A_{k}^{L}\prod_{a=1}^{3}\prod_{j=1}^{N_{a}/2}h_{q,k}^{(a)}\,(u_{a}^{j})\right]\sqrt{\frac{\det G_{+}}{\det G_{-}}}.\label{eq:ov_Ansatz}
\end{equation}

This follows on the one hand from concrete investigations~\cite{Buhl-Mortensen:2015gfd,deLeeuw:2016umh,DeLeeuw:2018cal,DeLeeuw:2019ohp} and on the other hand from  a proposed iterative procedure
for the determination of such overlaps~\cite{Gombor:2020auk} which builds on a number of assumptions.  
The upper limit of the sum is equal to the bond dimension of the matrix product
state and in the iterative process it originates from the fact that one cuts open the trace appearing in the matrix product state and 
sums over diagonal contributions. The constants $A_k^L$ encode the overlap of the vacuum state of the initial homogeneous spin chain with the matrix product state. 
We define the open version of the matrix product state as
\begin{equation}
\langle\mathrm{MPS}_{q,\theta,j,k}|=\left(M_{12}^{q,\theta}\dots M_{2L-1,2L}^{q,\theta}\right)_{j,k},
\end{equation}
and choosing a concrete parametrization of the $SU(2)$ generators we compute
\begin{equation}
A_{k}^{L}=\langle\mathrm{MPS}_{q,\theta,k,k}|0\rangle=\left(\frac{\sin(2\theta)}{2}k\right)^{L},
\end{equation}
where $|0\rangle$ is the state corresponding to the operator~\eqref{vac}.
The remaining factors in the overlap formula depend on the Bethe roots of the eigenstate in question and trivialize
to unity for the vacuum state.  In particular, $G$ is the Gaudin matrix whose determinant factorizes
as $\det G=\det G_+ \det G_-$  for Bethe states with paired root configurations~\cite{Gaudin:1976sv}.
For the alternating $SU(4)$ spin chain of ABJM theory there are three types of Bethe roots  of
which two are momentum carrying~\cite{Minahan:2008hf}. The $h$-factors in the overlap formula can be found by an iterative
procedure that relies on the nested Bethe ansatz. The procedure builds on the assumption that at each level of nesting the
boundary state can be written as a two-site product state expressed in terms of a reflection matrix $K_{ab}(p)$
that is given as the asymptotic overlap between the boundary state and a two-particle state with particle momenta $p$ and $-p$ and quantum numbers $a$ and $b$. For each level of nesting the overlap formula picks up a reflection factor corresponding
to the vacuum state at that level and at the last level one is left with an overlap of a $SU(2)$ spin chain which  
is well-known. 

The nested Bethe ansatz in the present case involves two levels of nesting. At the first level, at the odd sites of the spin chain, one replaces a field of type $Y_1$ with an excitation in the form of a field of type $Y_2$ or $Y_3$. Similarly, at the even sites of the spin chain one
replaces a field of type $Y_4^{\dagger}$ with a field of type $Y_2^{\dagger}$ or $Y_3^{\dagger}$. Level one excitations at the
odd sites carry Bethe rapidities $u_1^j$, $j=1,\ldots, N_1$ and level one rapidities at the even sites carry Bethe rapidities $u_3^k,$ $k=1,\ldots, N_3$.  At the second level, where there are four different neighbor configurations, 
one effectively has an inhomogeneous $SU(2)$ spin chain of length $N_1+N_3$ and inhomogeneities $u_1^j$, $u_3^k$. The excitations of this chain carry rapidities 
$u_2^i$, $i=1,\ldots, N_2$.  As argued in~\cite{Kristjansen:2021abc} the selection rule for a Bethe eigenstate to have a non-vanishing overlap with the matrix product states under consideration is the following type
of pairing of roots 
$$\{u_1^j\}_{j=1}^{N_1}=\{-u_3^k\}_{k=1}^{N_3},\hspace{0.5cm} \{u_2^j\}_{j=1}^{N_2}=\{-u_2^j\}_{j=1}^{N_2},$$
where we note that in particular $N_3=N_1$.
The achiral pair structure is also a consequence of the untwisted integrability condition \eqref{eq:intcond}, see the argument in \cite{Gombor:2020kgu,Gombor:2021hmj}. 

In the following we use the mapping of local operators to spin chain states
\begin{equation}
 Y^{A_1}Y^\dagger_{A_2}\dots Y^{A_{2L-1}}Y^\dagger_{A_{2L}} \to 
 |A_1,A_2,\dots,A_{2L-1},A_{2L}\rangle, 
\end{equation}
therefore the pseudo-vacuum is
\begin{equation}
 |0\rangle = |1,4,1,4,\dots,1,4\rangle.
\end{equation}
We also define the $\mathfrak{gl}_4$ generators $(E_{A,B}){}_{n}$ which perform the state replacement $A\to B$ at site $n$. 
At the first level of the nested coordinate Bethe Ansatz we can write the two-particle Bethe state as
\begin{eqnarray}
\lefteqn{ \hspace*{-.5cm}|u_{1},u_{3}\rangle_{a,b}= } \\
&& \hspace*{-1.2cm}\sum_{n,m=1}^{L}e^{ip^{1}n+ip^{3}m}\sum_{c,d=1}^{2}\chi_{a,b}^{c,d}(u_{1}-u_{3})|2n-1,2m\rangle_{c,d}\nonumber \\
&& \hspace*{-1.4cm} +\sum_{n=1}^{L}e^{i(p^{1}+p^{3})n}\left(\zeta_{a,b}^{(1)}(u_{1},u_{3})|2n-1\rangle+\zeta_{a,b}^{(2)}(u_{1},
u_{3})|2n\rangle\right), \nonumber
\end{eqnarray}
where $a,b,c,d=1,2$ with  $a$ and $c$ labelling the two types of excitations on the odd sites and $b$ and $d$ labelling the two types of excitations on the even sites. The notation should be almost self-explanatory; when we replace the $n$'th $Y^1$ and $m$'th $Y^\dagger_4$ of the pseudo-vacuum by $Y^{c+1}$ and $Y^\dagger_{4-d}$ we define the 
states as
\begin{equation}\nonumber
|2n-1,2m\rangle_{c,d}  =(-1)^{d-1}(E_{c+1,1}){}_{2n-1}(E_{4-d,4}){}_{2m}|0\rangle,
\end{equation}
and when we replace a $Y^1$ or a $Y^\dagger_4$ of the pseudovacuum to $Y^{4}$ or $Y^\dagger_{1}$ we define the 
states as
\begin{equation}\nonumber
|2n-1\rangle  =(E_{4,1}){}_{2n-1}|0\rangle,\qquad|2n\rangle=(E_{1,4}){}_{2n}|0\rangle.
\end{equation}
The expansion coefficients are found to be 
\begin{equation} \nonumber
\chi_{a,b}^{c,d}(u)=\begin{cases}
\delta_{a,b}^{c,d}, & 2n-1<2m\\
R_{a,b}^{c,d}(u), & 2n-1>2m
\end{cases}
\end{equation}
where
\begin{equation} \nonumber
R_{a,b}^{c,d}(u)=\frac{u\delta_{a}^{c}\delta_{b}^{d}-i\delta_{a}^{d}\delta_{b}^{c}}{u-i},
\end{equation}
and 
\begin{equation} \nonumber
\zeta_{a,b}^{(1)}(u,v)=\epsilon_{a,b}\frac{-v+i/2}{u-v-i},\hspace{0.5cm}\zeta_{a,b}^{(2)}(u,v)=\epsilon_{a,b}\frac{u+i/2}{u-v-i}.
\end{equation}
Finally, the momentum is related to the rapidity as
\begin{equation}
e^{ip}=\frac{u+i/2}{u-i/2}.
\end{equation}
From the concrete expression for the two-particle state at the first level of nesting we can find the  expression 
for the reflection factor at this level
\begin{eqnarray}\label{SU2matrix}
\lefteqn{\hspace*{-0.6cm}K_{a,b}^{(q,k)}(u)=  \lim_{L\to\infty}\frac{A_{k}^{-L}\langle\mathrm{MPS}_{q,\theta,k,k}|u,-u\rangle_{a,b}}{L}=}
\\
& & \hspace*{-0.6cm}\frac{(u+i/2)(u+i(q-\frac{1}{2}))}{(u+i(k-\frac{1}{2}))(u+i(k+\frac{1}{2}))} \times \nonumber \\
&& \hspace*{-1.5cm}
 \left(\begin{array}{cc}
1 & -\cot(2\theta)+\frac{i}{\sin(2\theta)}\frac{k}{u-i/2}\\
-\cot(2\theta)-\frac{i}{\sin(2\theta)}\frac{k}{u-i/2} & -1
\end{array}\right). \nonumber
\end{eqnarray}
As announced earlier, at the second level of nesting we now effectively have a reflection matrix characteristic
of a $SU(2)$ XXX spin chain whose vacuum has a non-trivial overlap with the matrix product state.
Choosing the vacuum at this level of nesting to be the 1-state the pre-factor of the
matrix in eqn.~\eqref{SU2matrix} encodes the overlap of the vacuum state with the matrix product state, i.e.
\begin{equation}
h_{q,k}^{(1)}(u)=K_{1,1}^{(q,k)}(u)=\frac{(u+i/2)(u+i(q-\frac{1}{2}))}{(u+i(k-\frac{1}{2}))(u+i(k+\frac{1}{2}))} ,
\end{equation}
whereas the matrix itself leads to the following overlap function~\cite{Pozsgay:2018ixm}
\begin{equation}
h_{q,k}^{(2)}(u)=\frac{1}{\sin(2\theta)^{2}}\frac{u^{2}+k^{2}}{u\sqrt{u^{2}+1/4}}.
\end{equation}
Under the assumption that the matrix product state decomposes into two-site product states at both levels of nesting the overlap will receive $(N_1+N_3)/2=N_1$ factors of type
$h_{q,k}^{(1)}(u)$ and $N_2/2$ factors of $h_{q,k}^{(2)}(u)$.

Summarizing, we get for the complete overlap formula
\begin{align}
\frac{\langle\mathrm{MPS}_{q,\theta}|\mathbf{u}\rangle}{\sqrt{\langle\mathbf{u}|\mathbf{u}\rangle}}=
\left({\sin(2\theta)}\right)^{L}\left(\frac{1}{\sin(2\theta)}\right)^{N_{2}}    
\frac{\langle\mathrm{MPS}_{q}|\mathbf{u}\rangle}{\sqrt{\langle\mathbf{u}|\mathbf{u}\rangle}},
\end{align}
where all $\theta$ dependence is explicitly exposed. 
We are interested in the limit $\theta\rightarrow 0$ and we immediately  see
that the requirement of a non-vanishing result imposes the selection rule $N_2=L$ which is in accordance with the analysis carried
out for $q=2$ in~\cite{Gombor:2021hmj,Kristjansen:2021abc}. 
We note, however, that the $\theta$ independent part of the overlap is well-defined not only for $N_2=L$ but for any Bethe state with
a paired root configuration. More precisely, it reads in terms of Baxter polynomials
\begin{align}
\frac{\langle\mathrm{MPS}_{q}|\mathbf{u}\rangle}{\sqrt{\langle\mathbf{u}|\mathbf{u}\rangle}}=
&\sum_{k=1} ^{q-1}
\frac{Q_{1}(-\frac{i}{2})Q_{1}(-i(q-\frac{1}{2}))}{Q_{1}(-i(k-\frac{1}{2}))Q_{1}(-i(k+\frac{1}{2}))}\,\, \, \times \nonumber\\
&\left(\frac{k}{2}\right)^L\!\!\!\frac{Q_{2}(-ik)}{\sqrt{\bar{Q}_{2}(0)\bar{Q}_{2}(-i/2)}}\sqrt{\frac{\det G_{+}}{\det G_{-}}}.\label{eq:general_form}
\end{align}
Here we have introduced $\bar{Q}_2(u)$ which is the Baxter polynomial $Q_2$ with  possible zero roots excluded. The rationale behind this kind of regularization 
was explained in~\cite{Kristjansen:2021xno}.
We notice that for $q=2$ the sum reduces to a single term corresponding 
to $k=1$ for which the first line of the expression becomes equal to one.  Thus, the  $q=2$ result of~~\cite{Gombor:2021hmj} is correctly recovered. 

\section{Validation\label{Validation}}

\begin{table}
\begin{centering}
\begin{tabular}{|c|c|c|c|}
\hline 
$(N_{1},N_{1},N_{3})$ & $Q_{1}(u)$ & $Q_{2}(u)$ & $Q_{3}(u)$\tabularnewline
\hline 
\hline 
$(0,0,0)$ & \textcolor{red}{$1$} & \textcolor{red}{$1$} & \textcolor{red}{$1$}\tabularnewline
\hline 
$(1,0,0)$ & $u$ & $1$ & $1$\tabularnewline
\hline 
$(0,0,1)$ & $1$ & $1$ & $u$\tabularnewline
\hline 
$(1,0,1)$ & \textcolor{red}{$u$} & \textcolor{red}{$1$} & \textcolor{red}{$u$}\tabularnewline
\hline 
\multirow{4}{*}{$(1,1,1)$} & $u+\frac{1}{2}$ & $u+\frac{1}{2}$ & $u+\frac{1}{2}$\tabularnewline
\cline{2-4} \cline{3-4} \cline{4-4} 
 & $u-\frac{1}{2}$ & $u-\frac{1}{2}$ & $u-\frac{1}{2}$\tabularnewline
\cline{2-4} \cline{3-4} \cline{4-4} 
 & \textcolor{red}{$u+\frac{1}{2\sqrt{3}}$} & \textcolor{red}{$u$} & \textcolor{red}{$u-\frac{1}{2\sqrt{3}}$}\tabularnewline
\cline{2-4} \cline{3-4} \cline{4-4} 
 & \textcolor{red}{$u-\frac{1}{2\sqrt{3}}$} & \textcolor{red}{$u$} & \textcolor{red}{$u+\frac{1}{2\sqrt{3}}$}\tabularnewline
\hline 
\multirow{2}{*}{$(2,2,2)$} & \textcolor{red}{$u^{2}-\frac{3}{20}$} & \textcolor{red}{$u^{2}-\frac{1}{5}$} & \textcolor{red}{$u^{2}-\frac{3}{20}$}\tabularnewline
\cline{2-4} \cline{3-4} \cline{4-4} 
 & \textcolor{red}{$u^{2}+\frac{1}{4}$} & \textcolor{red}{$u^{2}+\frac{1}{3}$} & \textcolor{red}{$u^{2}+\frac{1}{4}$}\tabularnewline
\hline 
\end{tabular}
\par\end{centering}
\caption{All $Q$-functions for $L=2$. The red solutions have pair structure. }

\label{tab:L2}
\end{table}

\begin{table*}[t]
\begin{centering}
\begin{tabular}{|c|c|c|}
\hline 
$(N_{1},N_{2},N_{3})$ & $Q_{1}(u)=(-1)^{N_{1}}Q_{3}(-u)$ & $Q_{2}(u)$ \tabularnewline
\hline 
\hline 
$(0,0,0)$ & $1$ & $1$\tabularnewline
\hline 
\multirow{2}{*}{$(1,0,1)$} & $u+\frac{1}{2\sqrt{3}}$ & $1$\tabularnewline
\cline{2-3} \cline{3-3} 
 & $u-\frac{1}{2\sqrt{3}}$ & $1$\tabularnewline
\hline 
\multirow{3}{*}{$(1,1,1)$} & $u$ & $u$\tabularnewline
\cline{2-3} \cline{3-3} 
 & $u+\frac{1}{2}$ & $u$\tabularnewline
\cline{2-3} \cline{3-3} 
 & $u-\frac{1}{2}$ & $u$\tabularnewline
\hline 
\multirow{1}{*}{$(2,1,2)$} & $u^{2}-\frac{1}{12}$ & $u$\tabularnewline
\hline 
\multirow{6}{*}{$(2,2,2)$} & $u^{2}-\frac{1}{28}(9+4\sqrt{2})$ & $u^{2}-\frac{1}{7}(1+2\sqrt{2})$\tabularnewline
\cline{2-3} \cline{3-3} 
 & $u^{2}-\frac{1}{28}(9-4\sqrt{2})$ & $u^{2}-\frac{1}{7}(1-2\sqrt{2})$\tabularnewline
\cline{2-3} \cline{3-3} 
 & $u^{2}-\frac{1}{12}\sqrt{\frac{1}{2}\left(67-\sqrt{73}\right)}u+\frac{1}{48}(7-\sqrt{73})$ & $u^{2}+\frac{1}{32}(3-\sqrt{73})$\tabularnewline
\cline{2-3} \cline{3-3} 
 & $u^{2}+\frac{1}{12}\sqrt{\frac{1}{2}\left(67-\sqrt{73}\right)}u+\frac{1}{48}(7-\sqrt{73})$ & $u^{2}+\frac{1}{32}(3-\sqrt{73})$\tabularnewline
\cline{2-3} \cline{3-3} 
 & $u^{2}-\frac{1}{12}\sqrt{\frac{1}{2}\left(67+\sqrt{73}\right)}u+\frac{1}{48}(7+\sqrt{73})$ & $u^{2}+\frac{1}{32}(3+\sqrt{73})$\tabularnewline
\cline{2-3} \cline{3-3} 
 & $u^{2}+\frac{1}{12}\sqrt{\frac{1}{2}\left(67+\sqrt{73}\right)}u+\frac{1}{48}(7+\sqrt{73})$ & $u^{2}+\frac{1}{32}(3+\sqrt{73})$\tabularnewline
\hline 
\multirow{4}{*}{$(3,3,3)$} & $u^{3}+\left(\frac{1}{4}-\sqrt{\frac{2}{5}}\right)u$ & $u^{3}+\frac{1}{15}\left(5-4\sqrt{10}\right)u$\tabularnewline
\cline{2-3} \cline{3-3} 
 & $u^{3}+\left(\frac{1}{4}+\sqrt{\frac{2}{5}}\right)u$ & $u^{3}+\frac{1}{15}\left(5+4\sqrt{10}\right)u$\tabularnewline
\cline{2-3} \cline{3-3} 
 & $u^{3}+\frac{3}{20}u+\frac{1}{20}\sqrt{\frac{13}{3}}$ & $u^{3}+\frac{1}{5}u$\tabularnewline
\cline{2-3} \cline{3-3} 
 & $u^{3}+\frac{3}{20}u-\frac{1}{20}\sqrt{\frac{13}{3}}$ & $u^{3}+\frac{1}{5}u$\tabularnewline
\hline 
\end{tabular}
\par\end{centering}
\caption{All $Q$-functions with pair structure for $L=3$.}

\label{tab:L3}
\end{table*}

The recursive derivation carried out above relies on the assumption that  the matrix product state has an asymptotic factorization into two site states at each level of
nesting~\cite{Gombor:2020auk}. Instead of engaging into formally proving this we will simply perform a detailed numerical test of our formula.
As~\eqref{eq:general_form} was proven analytically for $q=2$ in~\cite{Gombor:2021hmj} it suffices to validate the ratio
\begin{eqnarray}
\lefteqn{\frac{\langle\mathrm{MPS}_{q}|\mathbf{u}\rangle}{\langle\mathrm{MPS}_{2}|\mathbf{u}\rangle}=
\frac{\langle\mathrm{MPS}_{q,\theta}|\mathbf{u}\rangle}{\langle\mathrm{MPS}_{2,\theta}|\mathbf{u}\rangle}=}\\
&&\sum_{k=1}^{q-1}k^{L}\frac{Q_{1}(-\frac{i}{2})Q_{1}(-i(q-\frac{1}{2}))}{Q_{1}(-i(k-\frac{1}{2}))Q_{1}(-i(k+\frac{1}{2}))}\frac{Q_{2}(-ik)}{Q_{2}(-i)}. \nonumber
\end{eqnarray}
As already pointed out, the ratio is well-defined for any Bethe state with paired root configurations. We validated the formula for all Bethe states with paired root configurations
with $L=2,3$. 

The relevant Bethe eigenstates are most conveniently described by associating the following $\mathfrak{gl}(4)$ weights to the possible spin states at the odd, respectively even
sites of the spin chain
\begin{eqnarray}
Y^1, Y_1^\dagger \to (\pm 1,0,0,0),\quad Y^2, Y_2^\dagger \to (0\pm 1,0,0), \\
Y^3, Y_3^\dagger \to (0,0,\pm 1,0),\quad Y^4, Y_4^\dagger \to (0,0,0,\pm 1).
\end{eqnarray}
The $\mathfrak{gl}(4)$ weights of a Bethe state with root numbers
$N_{1},N_{2},N_{3}$ are then $(L-N_{1},N_{1}-N_{2},N_{2}-N_{3},N_{3}-L)$.
The full Hilbert space is $\left((1,0,0,0)\otimes(0,0,0,-1)\right)^{\otimes L}$.
The fusion rule for $L=2$ reads 
\begin{eqnarray}
\lefteqn{\hspace{-0.5cm}\left((1,0,0,0)\otimes(0,0,0,-1)\right)^{\otimes2}  =}\\
& &\hspace{-0.5cm} =(2,0,0,-2)\oplus(1,1,0,-2)\oplus(2,0,-1,-1) \oplus \nonumber \\
&&\hspace{-0.15cm}(1,1,-1,-1)\oplus4(1,0,0,-1)\oplus2(0,0,0,0), \nonumber
\end{eqnarray}
and for $L=3$ we obtain 
\begin{eqnarray}
\lefteqn{\hspace{-0.5cm}\left((1,0,0,0)\otimes(0,0,0,-1)\right)^{\otimes3}  =} \\
 &&\hspace{-0.5cm} =(3,0,0,-3)\oplus2(2,1,0,-3)\oplus2(3,0,-1,-2)\oplus \nonumber \\
 &&\hspace{-0.5cm} 4(2,1,-1,-2) \oplus 9(2,0,0,-2)\oplus 9(1,1,0,-2)\oplus \nonumber \\
&& \hspace{-0.5cm}  9(2,0,-1,-1)\oplus (1,1,1,-3)\oplus(3,-1,-1,-1)\oplus \nonumber \\
 &&\hspace{-0.5cm} 8(1,1,-1,-1)\oplus18(1,0,0,-1)\oplus6(0,0,0,0). \nonumber
\end{eqnarray}
From the explicit solutions of the Bethe equations we observe that there are 6 Bethe vectors with non-vanishing overlap for $L=2$ and 17 such vectors for $L=3$. We have checked the overlap formula for all of these vectors. The tables \ref{tab:L2} and \ref{tab:L3} summarize the corresponding $Q$-functions. The recursive strategy has once again proven its worth and this time in deriving a formula that had not been presented beforehand.

\section{Conclusion\label{Conclusion}}
We have found a closed formula for the overlaps of the Bethe eigenstates of the alternating integrable $SU(4)$ spin chain, describing the scalar operators of ABJM theory, and particular matrix product states of arbitrary bond dimension which represent co-dimension one 1/2 BPS domain walls in the field theory.  This demonstrates that the domain walls in question constitute integrable boundary states which is in agreement with the analysis of~\cite{Linardopoulos:2022wol} where a similar 
conclusion was  reached the  from the string theory side 
when considering a semi-classical limit corresponding to the limit of large bond dimension.
With the derivation of the overlap formula we have determined the value of the one-point functions of all scalar operators in the presence of particular domain walls, at least at the leading order in perturbation theory.  
An interesting question is whether the formula can be extended to the full field theory by exploiting the idea of covariance under fermionic duality transformations~\cite{Kristjansen:2020vbe},
as implemented for the matrix product state of bond dimension 2 in~\cite{Kristjansen:2021abc}.  It appears that the
arguments of~\cite{Kristjansen:2021abc} carry over to the present case since the $Q$-functions in the first line of 
eqn.\ \eqref{eq:general_form} are unaffected by the duality transformations considered.
It would likewise be interesting if the leading order result could be bootstrapped to higher loop orders as it was the case for ${\cal N}=4$ SYM
theory  and its string theory dual~\cite{Buhl-Mortensen:2017ind,Wang:2020seq,Komatsu:2020sup,Gombor:2020kgu,Gombor:2020auk}. In this connection, it might also be useful or necessary for  
consistency checks to set up the perturbative program for defect ABJM theory as it was done for the various versions of defect ${\cal N}=4$ SYM theory \cite{Buhl-Mortensen:2016pxs,Buhl-Mortensen:2016jqo,GimenezGrau:2018jyp,Gimenez-Grau:2019fld}. 
 
 Our novel overlap formula is the latest addition to the growing list of exact overlap formulas for integrable boundary states of various spin chains. Recently, a novel type of
 integrable
boundary states, cross-cap states~\cite{Caetano:2021dbh}, were identified in 1+1 dimensional field theories and in the rank one integrable  $SU(2)$ and $SL(2,R)$ spin chains.  Cross-cap states should also have an analogue for  nested spin chains and in particular for the alternating spin chain considered 
here with possible implications for ABJM theory and
its 
 dual
type IIA string theory. Finally, one could 
 imagine that the yet lower dimensional versions of the AdS/CFT correspondence have more integrable boundary states waiting to be discovered.

 \vspace{0.3cm}

\section*{Acknowledgments}
T.G.\ was supported by the NKFIH grant K134946. 
C.K.\ was supported by DFF-FNU through grant number 1026-00103B.  C.K. thanks K.\ Zarembo for
discussions.




  \bibliographystyle{elsarticle-num} 

 \bibliography{ABJM}





\end{document}